\documentclass[]{aastex631}
\usepackage{url}

\accepted{by ApJ, June 21, 2022}

\shorttitle{Universal Relations}
\shortauthors{Konstantinou \& Morsink}

\graphicspath{Figures/}
    \usepackage{graphicx}
    \usepackage{capt-of}
    \usepackage{lipsum}

\begin{document}

\title{Universal Relations for the Increase in the Mass and Radius of a Rotating Neutron Star}

\correspondingauthor{Sharon Morsink}
\email{sharon.morsink@ualberta.ca}

%\nocollaboration
\author[0000-0002-1072-7313]{Andreas Konstantinou}
\affiliation{Department of Physics\\
University of Alberta \\
Edmonton AB, T6G 2E1, Canada}

\author[0000-0003-4357-0575]{Sharon M. Morsink}
\affiliation{Department of Physics\\
University of Alberta \\
Edmonton AB, T6G 2E1, Canada}

\begin{abstract}

Rotation causes an increase in a neutron star's mass and equatorial radius. The mass and radius depend sensitively on the unknown equation of state (EOS) of cold, dense matter.  However, the increases in mass and radius due to rotation are almost independent of the EOS. The EOS independence leads to the idea of neutron star universality. In this paper, we compute sequences of rotating neutron stars with constant central density. We use a collection of randomly generated EOS to construct simple correction factors to the mass and radius computed from the equations of hydrostatic equilibrium for non-rotating neutron stars.  The correction factors depend only on the non-rotating star's mass and radius and are almost independent of the EOS.  This makes it computationally inexpensive to include observations of rotating neutron stars in EOS inference codes. We also construct a mapping from the measured mass and radius of a rotating neutron star to a corresponding non-rotating star. The mapping makes it possible to construct a zero-spin mass-radius curve if the masses and radii of many neutron stars with different spins are measured.

We show that the changes in polar and equatorial radii are symmetric, in that the polar radius shrinks at the same rate that the equatorial radius grows. This symmetry is related to the observation that the equatorial compactness (the ratio of mass to radius) is almost constant on one of the constant-density sequences.

\end{abstract}


\section{Introduction} \label{sec:intro}

The cores of neutron stars provide a unique combination of low temperature and high density that can provide insights about the equation of state (EOS) of supranuclear density matter \citep{2016Watts}.  While the EOS of the matter in a neutron star is still undetermined, each proposed EOS is mapped to a unique mass-radius curve by the relativistic equations of stellar structure. As a result, the measurement of the masses and radii of many neutron stars can, in principle, be used to determine the neutron star's EOS \citep{1992Lindblom,2012Lindblom,2014Lindblom}. This is one of the main motivations for the Neutron Star Interior Composition Explorer (NICER) X-ray telescope's observations of rotation-powered pulsars \citep{2012SPIE.8443E..13Gendreau}.

The use of mass and radius measurements of neutron stars to constrain the cold dense EOS is typically done through the framework of statistical inference (see, e.g. \cite{2020ApJ...888...12Miller}). A set of $N$ neutron stars, each with known spin are observed, and a likelihood function for each neutron star's mass and equatorial radius is found. These likelihood functions could come from pulse-profile modelling of x-ray timing data, as is done with NICER observations \citep{2019Miller,2019Riley,2021ApJ...918L..28Miller, 2021Riley}, for example. Alternatively, spectroscopic observations of neutron stars in quiescent low-mass x-ray binaries or thermonuclear bursts on the surfaces of accreting neutron star could yield likelihood 
functions \citep{2016Ozel}. Given a set of parameters describing an EOS, the relativistic stellar structure equations predict values of mass and equatorial radius for each neutron star's particular value of spin. Each neutron star is assumed to be described by the same set of EOS parameters, but each will have a different density at its centre. This allows the calculation of the likelihood that a particular set of EOS parameters correctly describes the set of $N$ neutron star observations.

Although the procedure described in the previous paragraph is well-defined, if many sets of EOS parameters are to be tested, the computation of rapidly rotating neutron star models would make this time-consuming and computationally expensive. In practice, the neutron stars observed by NICER are rotating slowly, with spin frequencies near $200$ Hz. As a result, the computation of rotating models is not necessary, since the masses and radii do not change drastically at slow rotation rates. EOS inference for the NICER neutron stars makes use of the spherically symmetric TOV equations \citep{2019Miller,2019Raaijmakers,2021ApJ...918L..28Miller,2021ApJ...918L..29Raaijmakers} to map between EOS parameters and the mass-radius values. While the use of the TOV equations does not introduce any bias in the NICER results, future observations, either with better precision, or of more rapidly rotating neutron stars could lead to biased results. The goal of this paper is to evaluate the magnitude of the inaccuracies that result from the TOV approximation in EOS inference, and to provide a simple, computationally inexpensive method for including rotation in EOS inference codes.

The concept of universality is useful when considering the effects of rotation on neutron stars. Black holes are truly universal: specifying only the mass and dimensionless spin parameter leads to a complete description of the black hole's gravitational field and properties. Neutron stars, at first glance, do not seem to have universal properties. Each proposed EOS predicts a different curve of mass as a function of radius, $M(R)$. However, other properties, such as the moment of inertia \citep{1994ApJ...424..846Ravenhall} and  the gravitational binding energy \citep{2001ApJ...550..426Lattimer}, have an approximately universal dependence on the compactness, $C=M/R$, independent of the EOS.  When the neutron star rotates at angular velocity, $\Omega$, many properties depend approximately on only the dimensionless angular velocity, defined by $\bar{\Omega}^2 = \Omega^2 GM/R_e^3$, where $R_e$ is the equatorial radius, along with the compactness. For instance, approximately universal relations depending on $\bar{\Omega}$ and $C_e = M/R_e$ for the oblate shape of the neutron star \citep{2007Morsink,2021PhRvD.103f3038Silva} and the effective acceleration due to gravity over the neutron star's surface \citep{2014AlGendy} have been used to determine the equatorial radii of neutron stars by NICER \citep{2019bBogdanov}. {An approximately universal relation between the maximum mass rotating neutron star and the corresponding maximum mass for a non-rotating star \citep{1996ApJ...456..300Lasota,2016MNRAS.459..646Breu} has proven to be useful for placing limits on the properties of the remnants of neutron star mergers such as GW170817 \citep{2018ApJ...852L..25Rezzolla}.}
The oscillation modes of rotating neutron stars also exhibit universal properties {\citep{2020PhRvL.125k1106Kruger}, which could be used to constrain the EOS \citep{PhysRevD.105.124071Volkel}}.
An even stronger universality is the ``I-Love-Q" relation between the moments of inertia, Love numbers, and quadrupole moments of rotating neutron stars \citep{2013Yagi}. A more detailed review of neutron star universality is provided by \citet{2017PhR...681....1Yagi}.

In this paper we explore the universality of the {\em{change}} in the mass and radius of {uniformly} rotating neutron stars in order to introduce some approximate relations for the changes that occur when a neutron star spins. These increases are compared to non-rotating neutron stars with the same central density. The increases in mass and radius are written as simple expressions, independent of the EOS, depending only on the mass and radius of the non-rotating reference star as well as the angular velocity of the rotating star. This allows the construction of a set of simple correction factors for the masses and radii of neutron stars calculated using the spherically symmetric TOV equations. We also construct an inverse mapping from rotating stars to non-rotating stars to allow the reconstruction of the zero-spin mass-radius curve from observations of rotating neutron stars.

The structure of this paper is as follows. In Section \ref{sec:sequences} we introduce our library of randomly generated EOS, and compute sequences of rapidly rotating neutron stars connected by a constant value of central density. In Section \ref{sec:universal} we explore the dependence of the Kepler frequency on the mass and radius of the non-rotating member of a sequence, which we exploit to construct universal formulae for the increases in mass and radius due to rotation. In this section we also discuss a simple physical model for the universal behaviour seen and its application to EOS inference using only the solutions to the TOV equations. In Section \ref{sec:inverse} we construct a mapping from the measured properties of a rotating star to the properties of non-rotating neutron stars as a method to determine the zero-spin mass-radius curve.

%%%

\section{Spin-up sequences}
\label{sec:sequences}

We use the computer program \textit{Rapidly Rotating Neutron Star}, \texttt{rns},
\citep{1995Stergioulas} to solve the relativistic stellar structure equations for an axisymmetric, rotating neutron star given a zero-temperature EOS. The \texttt{rns} program is an adaptation of the Green function method \citep{1989Komatsu,1992Cook} for solving the Einstein field equations for axisymmetric stars, which is based on the Newtonian self-consistent field method \citep{1986Hachisu}.

\subsection{Equations of state}\label{sec:eos}

We created two separate families of randomly generated EOS. The first family is the piecewise polytrope (PP) \citep{Read2009}, constructed with the same range of parameters given by \citet{hlps-paper}. The second family of EOS is constructed using the speed-of-sound (CS) parametrization introduced by \citet{2019MNRAS.485.5363Greif}. For both families we randomly chose values for the various parameters using flat priors for the ranges specified in the respective papers. Initially we constructed separate universal relations for each family of EOS in order to test if there is any bias introduced by the choice of EOS family. However, the differences were minor, so we present results in this paper based on all of the randomly generated EOS from both families. We determined the optimal number of EOS to generate by first choosing a small number of EOS, computing the full set of rotating neutron stars, and then fitting the increase in radius caused by rotation. We then iteratively generated a new EOS and then fit the radius increase again, checking the convergence of the best-fit coefficients. We found that a set of 32 randomly generated EOS provided a fit to the mass that converged. The CS method creates a more restricted range of radii, so we limited our choices to 13 EOS from this family and used 19 EOS from the PP family in order to find empirical relations that are valid for a wider range of possible EOS.

The maximum masses (of non-rotating stars) generated from these 32 EOS range from $1.97 M_\odot$ to $2.88 M_\odot$. The radius of spherical 1.4 $M_\odot$ stars computed with the PP EOS range from 10.9 - 14.7 km, while the CS EOS stars range from 11.2 - 12.4 km.

Other methods for constructing phenomenological parametric EOS exist, such as the 
spectral method \citep{2014Lindblom}. \citet{2022PhRvD.105d3016Legred} have shown that parametric EOS can introduce correlations between the pressures at different densities that add artificial restrictions to the EOS that can bias inference results, and show that non-parametric methods, such as the Gaussian process method \citep{2019PhRvD..99h4049Landry} do not suffer from this type of bias. 

It is beyond the scope of this paper to explore all other types of phenomenological EOS. Instead we test the empirical relations derived in the following section against a small set of tabulated EOS to look for extreme EOS models where our results might fail. We included the following hadronic EOS: the APR EOS \citep{1998Akmal} includes 3-nucleon interactions and special relativistic corrections; the two BBB EOS \citep{1997A&A...328..274Baldo} as hadronic examples that have maximum masses smaller than $2 M_\odot$; an example of an EOS that includes hyperons (Table 5.8 of \citet{1997csnp.book.....Glendenning}), with a maximum mass below $1.6 M_\odot$; the ultra stiff EOS L \citep{1975Pandharipande} with a pion condensate. We also include three quark/hadron hybrid EOS \citep{2005Alford,2021arXiv211111919Kojo}, and a bare quark star \citep{1986ApJ...310..261Alcock}.

\subsection{Constant central-density spin-up sequences for one EOS}

In our application, we compute sequences of stars with the same EOS and central-density. The first star of the sequence is non-rotating, with a mass and radius denoted by $M_*$ and $R_*$ given by the solution of the TOV equations. The subsequent stars in the same sequence have the same central density, but increasing eccentricity. The values of mass and equatorial radius for each subsequent star in the sequence, $M$ and $R_e$, are larger than the corresponding values of $M_*$ and $R_*$ for the same central density. 

\begin{figure}

\plottwo{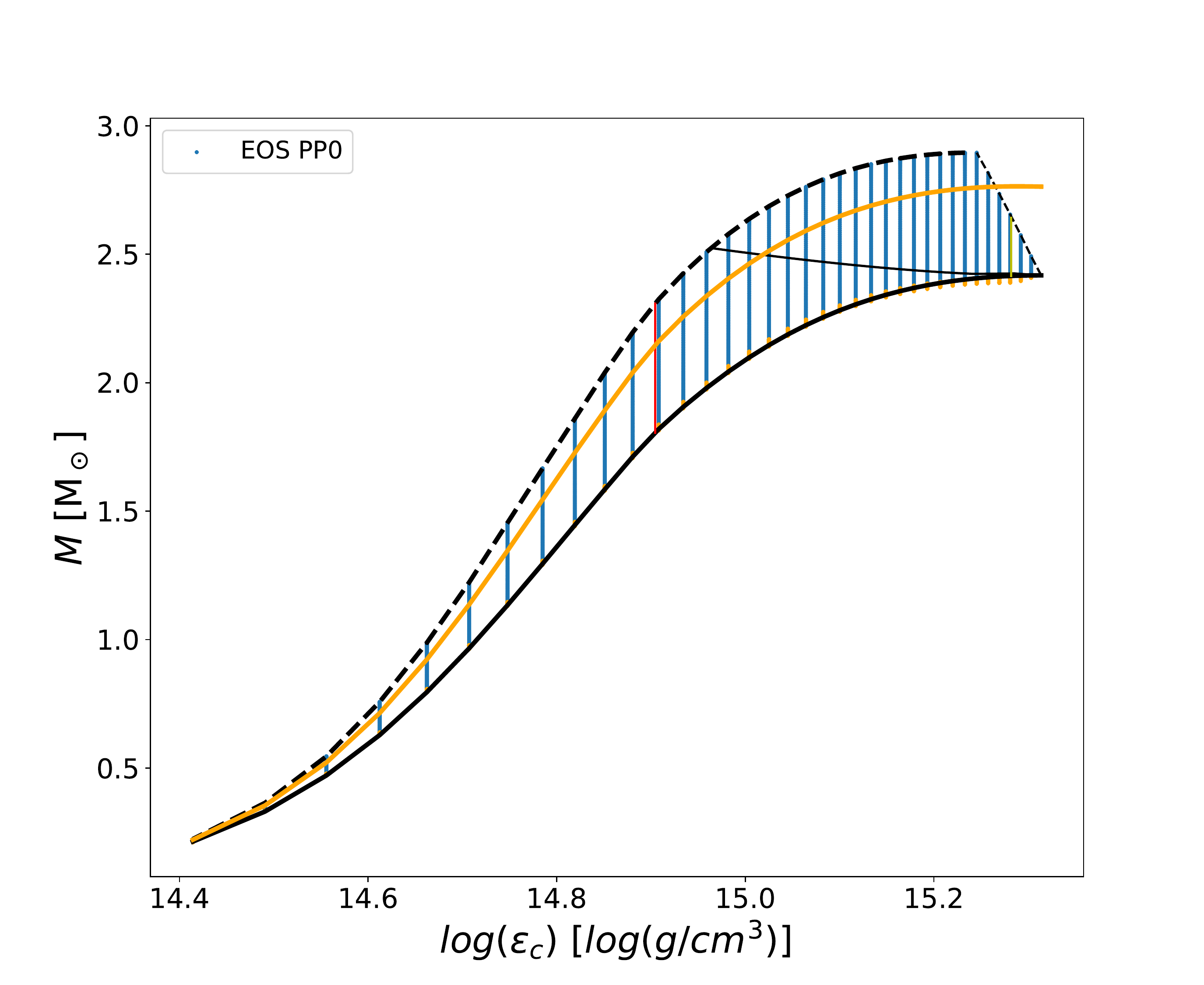}{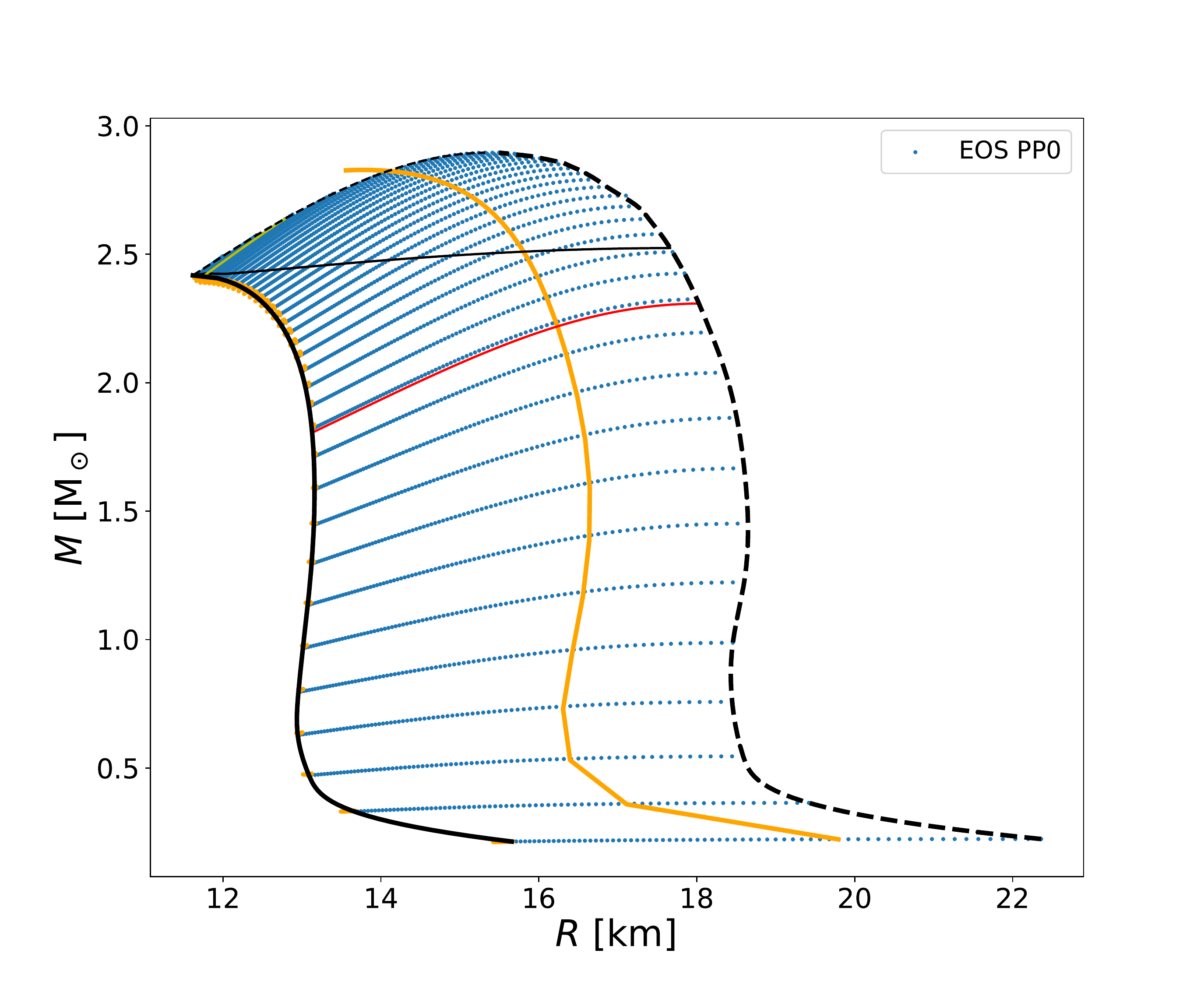}

    \caption{A set of constant central-density spin-up sequences for EOS PP0 showing mass versus central density (left) and mass versus equatorial radius (right).
    Each blue point represents the mass, equatorial radius, and central energy density of a rotating neutron star constructed with this EOS.  }
    \label{fig:m_r_nu_Pol0}
\end{figure}

As an example we provide a full set of constant central-density spin-up sequences for one sample EOS, the piecewise polytrope PP0, in Figure \ref{fig:m_r_nu_Pol0}. The choice of EOS to illustrate the sequences is not important, since the general features and shapes of the plots for each EOS are similar, even if the particular values of mass and radius are different. Each stable stellar model computed is represented as a point on these plots. Figure \ref{fig:m_r_nu_Pol0} (left pane) shows each spin-up sequence as a vertical series of dots. The lowest mass star in each sequence corresponds to a solution of the TOV equations. Solutions of the TOV equations are shown with a bold black curve.
s{The star with the largest mass on this curve is the TOV limit star.}
Stars with larger mass and same the central density in a sequence correspond to a star with a larger spin frequency. The two coloured vertical lines show the minimum central densities for the two densest polytropic regions for this particular PP. 

Each spin-up sequence terminates at the highest mass shown, which is also the highest frequency possible for a stable star in the sequence. The termination of the sequence is either due to the Kepler limit (also known as the mass-shed limit), shown with a bold dashed line, or the onset of the quasi-radial instability, shown with a thin dashed line. In the case of the maximum mass non-rotating star, there is only one member of this sequence, since a rotating star with the same central density as the maximum mass non-rotating star is unstable to quasi-radial modes. In our calculations we approximate the quasi-radial instability curve as a straight line in mass versus the logarithm of the central density as suggested by the results of \citet{1994Cook} (see their Figure 1, for example). While this is not exact, our results don't depend on the exact determination of the quasi-radial instability onset. 

This procedure allows us to compute a set of sequences of rotating neutron stars that include both normal and supramassive neutron stars \citep{1992Cook}. A solid curve dividing the models into normal and supramassive stars is shown in Figure \ref{fig:m_r_nu_Pol0}a, corresponding to stars with the same baryon mass as the maximum mass non-rotating stellar model. Stars below this curve are normal, while stars above the curve are supramassive.

The same constant-density spin-up sequences are shown in the right-hand pane of Figure \ref{fig:m_r_nu_Pol0}, where the mass and equatorial radius of each stable star are shown, along with the same curves described for the left-hand pane of Figure \ref{fig:m_r_nu_Pol0}. Each constant-density sequence is approximately a straight line with positive slope, so that the star's equatorial compactness is approximately constant as it spins faster. We will return to this property later in Section \ref{sec:compactness}. Additionally we add a gold curve, representing stars that are rotating at 95\% of the Kepler frequency, and gold points that overlap the TOV curve, representing an inverse map which will be explained in Section \ref{sec:inverse}.

Note that the sequences shown in Figure \ref{fig:m_r_nu_Pol0} are different from the constant baryon-mass sequences used in many applications (eg. \cite{1994Cook}) to model the spin-down of an isolated star that doesn't lose or accrete matter. In particular, it may be useful to compare the plots shown in Figure \ref{fig:m_r_nu_Pol0} with Figures 1 and 2 of \cite{1994Cook} (for a different EOS). In this paper, we are investigating the size of spin corrections to the mass and radius obtained from the TOV equations, so a comparison between models with the same central energy density is more useful. 

\section{Universal relations for the change in mass and radius}
\label{sec:universal}

We now investigate some approximately universal properties of neutron stars as they spin. Our approach is to consider dimensionless quantities, such as $M/M_*$, the mass of a rotating star divided by the mass of a non-rotating star with the same central density, and investigate how this ratio depends on different quantities. One natural choice of independent variable is $C_* = \frac{M_*}{R_*}\frac{km}{M_\odot}$, the non-dimensional compactness of the non-rotating star in a sequence. {(Note that many authors define compactness as the ratio $GM/Rc^2$ which differs from this definition by a constant factor.)} Similarly we could introduce the star's angular velocity normalized by $\Omega_* =  \left(GM_*/R_*^3)\right)^{1/2}$ {or alternatively, the dimensionless spin parameter $\chi = cJ/GM^2$, where $J$ is the star's angular momentum.}
(We use the following notation: the star's spin frequency, as measured at infinity is $\nu$, and the angular velocity, $\Omega$, is $\Omega = 2 \pi \nu$.) However, we have found that the behaviour of quantities such as $M/M_*$ for multiple EOS depends more closely on how close the angular velocity is to the Kepler limit than {either} the quantity ${\Omega}/\Omega_*${or $\chi$}. For this reason, we begin our exploration of universality with the Kepler limit.

\subsection{The Kepler limit}
\label{sec:kepler}

\begin{figure}[h]
    \centering
    \includegraphics[width=90mm]{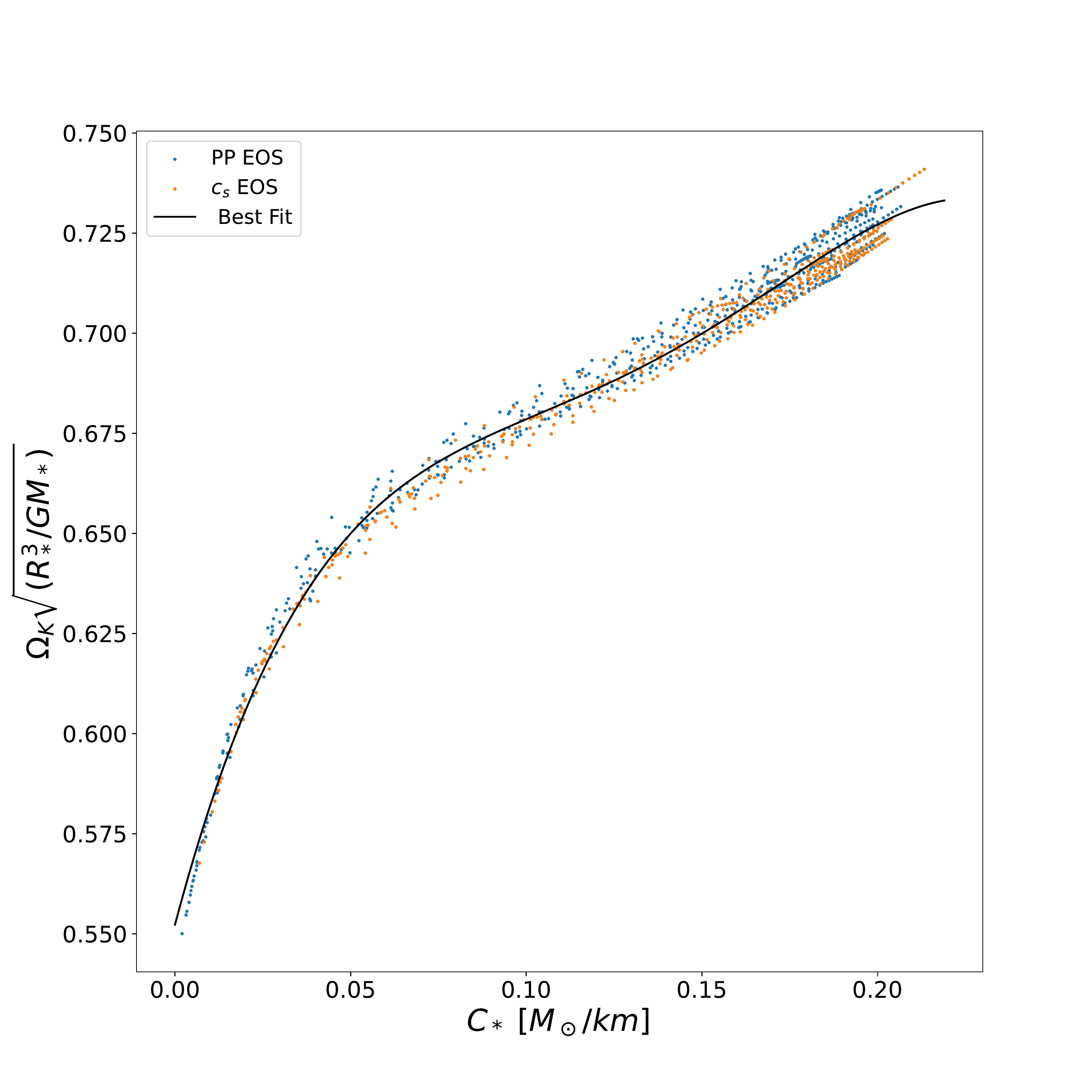}
    \caption{The Kepler angular velocity, $\Omega_K$ of neutron stars as a function of the compactness of the zero spin star with the same central density as the rotating star. Stars computed with a PP or CS EOS are shown with blue or gold dots. The best-fit curve, Equation (\ref{eq:OmegaK}) is shown as a solid curve. }
    \label{fig:kepler}
\end{figure}

The Kepler limit occurs when a star spins at the same rate that a test particle orbits at the star's equator. We use the symbol $\Omega_K$ to denote a star spinning at the Kepler limit. To investigate the Kepler limit, we first choose an EOS, and find all of the stellar models that are spinning at the Kepler limit. For instance, in the case of EOS PP0, this would correspond to the set of 31 points that terminate the constant-density spin-up sequences on the heavy dashed curve shown in Figure \ref{fig:m_r_nu_Pol0}. Each of these spin-up sequences has different non-rotating reference values for $M_*$ and $R_*$. The angular velocity for each of these stars is then normalized by $\Omega_*$ and plotted versus the compactness of the zero spin member of its sequence. This procedure is then repeated for each of the EOS in our library. The result is shown in 
Figure \ref{fig:kepler}.

Each point shown in Figure \ref{fig:kepler} corresponds to a star, constructed with any of the EOS, that is spinning at the Kepler limit. The angular velocity of the star, normalized by $\Omega_*$ follows a monotonic trend when plotted versus the compactness $C_*$ of the non-rotating reference star. The best-fit curve for this approximate relation for the Kepler angular velocity is given by a 4th
order polynomial,
\begin{eqnarray}
\Omega_K =   \Omega_*  \times \sum_{i=0}^{4} \; a_i C_*^i , \label{eq:OmegaK}
\end{eqnarray}
where the coefficients $a_i$ and the $R^2$ statistic are shown in Table \ref{tab:coeff}. This best-fit curve and all other fits in this paper were made using the Python packages \url{ numpy.polyfit} \citep{2020Natur.585..357H} for polynomial fits and \url{ scipy.optimize.curve_fit} \citep{2020SciPy-NMeth} for more complicated functions. 

\begin{table}[]
    \centering
    \begin{tabular}{ccccccccccc}
Equation &  Symbol       & $a_0$ & $a_1$ & $a_2$ & $a_3$ & $a_4$ & $a_5$ & A & B & $R^2$  \\
\hline
 (\ref{eq:OmegaK}) & $\Omega_K$ & 0.552 & 3.304 &-35.211 &180.61 &-326.48 & - & - & - & 0.9895\\
 (\ref{eq:Rfit}) & $R_e/R_*$ & - &-15.496& 442.60& -4945.62&23458.06&-40544.25 &0.203& 0.1611& 0.9939\\
(\ref{eq:Mfit}) & $M/M_*$ &      -0.0160&3.123&-20.721&41.202&-6.464&-&1.127&-&0.9808\\
\hline
    &  & $b_0$ & $b_1$ & $b_2$ & $b_3$ & $b_4$ & $b_5$ &
    $A_2$ & $B_2$ & $R^2$ \\
    \hline
    (\ref{eq:kepler-inv}) & $\Omega_{K2}$ &
    0.9933 & 0.4183 & -2.328 & -17.907 & 79.377 & - & - & - & 0.9121 \\
    (\ref{eq:Rinvfit}) & $R_e/R_*$ & - & -8.8981 & 216.845 & -2298.43 & 10211.98 & -16587.60 & 0.39998 & -0.0058314 & 0.9980 \\
   (\ref{eq:Minvfit}) & $M/M_*$ & - & 1.765 & -10.985 & 11.069 & 34.996 & - & - & - & 0.9722 \\
   \hline
      & &  & $d_1$ & $d_2$ & $d_3$ & $d_4$ &  &
     &  &  \\
     \hline
   (\ref{eq:Minvfit}) & $M/M_*$ & &   1.0 &     -2.924 & 15.305 & -9.908 &  
        
    \end{tabular}
    \caption{Coefficients and the $R^2$ statistic  for the best-fit equations. }
    \label{tab:coeff}
\end{table}

Given a solution of the TOV equations, $M_*$ and $R_*$, the empirical relation (\ref{eq:OmegaK}) predicts the largest possible angular velocity of a star that has the same central density. We will examine the mass and radius of this star in a later section. However, this formula does not predict whether or not this star would instead belong to a sequence that terminates at the quasi-radial instability, so our Kepler frequency formula has some limitations. Typical values for the Kepler frequency are shown in Table \ref{tab:typicalvalues}. For realistic ranges of mass and radius, the maximum spin frequency for a neutron star ranges over a factor of 2, from about 900 to 1900 Hz.

\begin{table}[h]
    \centering
    \begin{tabular}{cccr|cccc|cccc|cccc}
    $M_*$ & $R_*$ & $C_*$ & $\Omega_K/2\pi$ 
    & $\nu$ & $\Omega_n$& $\frac{M-M_*}{M_*}$ & $\frac{R_e-R_*}{R_*}$ 
    & $\nu$ &  $\Omega_n$&$\frac{M-M_*}{M_*}$ & $\frac{R_e-R_*}{R_*}$ 
    & $\nu$ &  $\Omega_n$&$\frac{M-M_*}{M_*}$ & $\frac{R_e-R_*}{R_*}$ \\
    $M_\odot$ & km & $M_\odot/$km & Hz  
    & Hz & & &  
    & Hz & & &
    & Hz & & \\
    \hline
        1.3 & 13 & 0.100 & 957 &  200 & 0.209 &	0.007 & 0.007
        & 400 &	0.418 &	0.028 &	0.025 
        & 600 &	0.627 &	0.072 &	0.041 \\
        1.4 & 11 & 0.127 & 1295 & 200 &	0.154 &	0.003 & 0.004
        & 400 &	0.309 &	0.015 &	0.014 
        & 600 &	0.463 &	0.035 &	0.027 \\
        1.6 & 13 & 0.123 & 1075 & 200 &	0.186 &	0.005 & 0.005
        & 400 &	0.372 &	0.022 &	0.020 
        & 600 &	0.558 &	0.055 &	0.035 \\
        1.8 & 12 & 0.150 & 1309 & 200 &	0.153 &	0.003 & 0.003
        & 400 &	0.305 &	0.013 &	0.013 
        & 600 &	0.458 &	0.033 &	0.025 \\
        2.0 & 10 & 0.200 & 1885 & 200 &	0.106 &	0.001 & 0.001
        & 400 &	0.212 &	0.005 &	0.005 
        & 600 &	0.318 &	0.012 &	0.011 

    \end{tabular}
    \caption{Spin corrections for typical neutron star parameters.}
    \label{tab:typicalvalues}
\end{table}

The empirical relation for the Kepler frequency can be compared with earlier calculations.
\citet{1989Natur.340..617Haensel} and \citet{1989PhRvL..62.3015Friedman} calculated the maximum angular velocity, $\Omega_{max}$ for a particular EOS, compared with the
mass, $M_{TOV}$ and radius, $R_{TOV}$ of the maximum mass non-rotating star with the same EOS, and found an approximately universal scaling of $\Omega_{max} = K (GM_{TOV}/R_{TOV}^3)^{1/2}$. \citet{1996ApJ...456..300Lasota} generalized this relationship so that $K$ is a function of the maximum compactness, $C_{TOV}$ of the non-rotating star. A somewhat different approach is to compare the maximum rotation rate of normal neutron stars with the non-rotating
neutron star with the same baryonic mass, as was done by \citet{1994Cook} (see their Figure 27). This mapping using the star's baryon mass does not yield a relation that appears as universal as the mapping using the central density shown in Figure \ref{fig:kepler}. Similar universal relations for the maximum rotation rates of neutron stars have been explored by a number of authors (e.g. \citet{2009A&A...502..605Haensel}, 
\citet{2018A&A...620A..69Haskell}, and
\citet{2020PhRvC.101a5805Koliogiannis}). We have verified that our results for the Kepler frequency agree (within reasonable error) with earlier work.

\subsection{Equatorial compactness ratio}
\label{sec:compactness}

The shape of a rotating neutron star is oblate, with an approximately universal shape \citep{2007Morsink} that depends on the equatorial compactness $C_e = \frac{M}{R_e} \frac{km}{M_\odot}$ and the angular velocity of the rotating star.
The hot spots on a rotation-powered X-ray pulsar can be at any latitude of the neutron star, but the universal relations for the shape allow the determination of the equatorial values of radius and compactness,  through pulse profile modelling \citep{2019bBogdanov} using X-ray timing telescopes such as NICER. 

It has been noted elsewhere \citep{2019Miller,2019Raaijmakers,2022PhRvX..12a1058Annala} that the equatorial compactness appears to be almost constant if the star's central density is kept constant as the star is spun up. This property can be seen, for example, for the case of slow rotation in Figure 5 of \citet{2019Raaijmakers}, for stars spinning at the Kepler limit in Figure 4 of \citet{2022PhRvX..12a1058Annala} {, and for a wide range of stellar rotation in Figure 3 of \citet{2022EPJC...82..512Mallick}}. 

In the example EOS PP0 shown in the right-hand panel of Figure \ref{fig:m_r_nu_Pol0}, the compactness is the slope of the curve formed by connecting the sequences of constant central density. The slopes of these sequences are approximately constant between the TOV limit (bold black curve) and the gold curve.
The gold curve connects neutron stars spinning at an angular velocity of 0.95 $\Omega_K$, as predicted by Equation (\ref{eq:OmegaK}). To the right of the gold curve the slope decreases somewhat as the sequences approach the Kepler limit, with departures from a constant slope that are more pronounced for less compact stars. This suggests that the equatorial compactness depends on both $C_*$ and the ``closeness"  of the star's rotation rate to the Kepler limit. We define a simple normalized angular velocity $\Omega_n$ of a particular rotating star
by
\begin{equation}
    \Omega_n(\Omega;C_*,\Omega_*) = \frac{\Omega}{\Omega_K(C_*,\Omega_*)},
\end{equation}
where $\Omega_K$ is given by the empirical fit shown in Equation (\ref{eq:OmegaK}). Each member of a constant-central density spin-up sequence is described by the same non-rotating reference values of $C_*$ and $\Omega_*$, so the normalized angular velocity $\Omega_n$ ranges from 0 to a value close to 1 for sequences that terminate at the Kepler limit. Note that due to the nature of the empirical fit, the highest spin star in any sequence could be slightly larger or smaller than 1. We use this same normalization for the sequences that terminate in the quasi-radial instability, which occurs at values of the normalized angular velocity that can be much smaller than 1. Typical values of $\Omega_n$ for realistic masses, radii, and spin frequencies are shown in Table \ref{tab:typicalvalues}.

\begin{figure}[h]
    \centering
    \includegraphics[width=90mm]{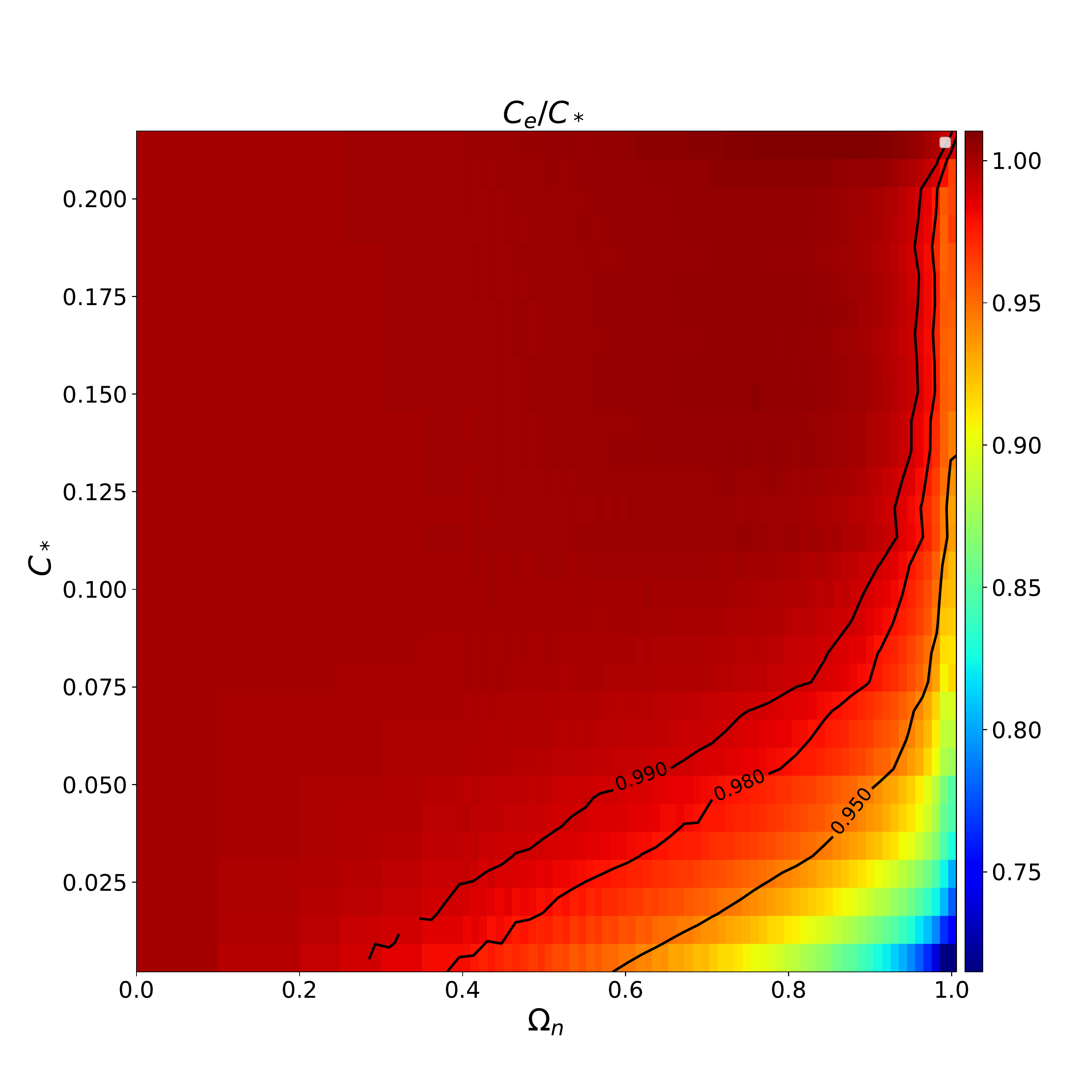}
    \caption{Average normalized equatorial compactness $C_e/C_*$ for rotating stars as a function of the compactness of the non-rotating star with the same central density and the normalized angular velocity. The colour in each bin is the average (over all 32 EOS) value of equatorial compactness of stars with the same value of $C_*$ and $\Omega_n$.  }
    \label{fig:compact2d}
\end{figure}

The dimensionless ratio of the equatorial compactness of a rotating neutron star to the compactness of a non-rotating star with same central density, $C_e/C_*$, is shown in Figure \ref{fig:compact2d}. This plot shows the dependence of $C_e/C_*$ on the normalized angular velocity $\Omega_n$ (horizontal axis) and the non-rotating compactness (vertical axis). The colour in each bin represents the average over all stars computed from the 32 EOS with the given values of $\Omega_n$ and $C_*$.  Although the ratio $C_e/C_*$ is not exactly 1.0, the contours show that for stars rotating at less than about 90\% of the Kepler frequency, the decrease in compactness (ie $1-C_e/C_*$) is less than 1\% for neutron stars with astrophysical masses and radii ($C_* \ge 0.1$). Although the stars with $C_* < 0.1$ have unrealistically low masses, these stars are included on the figure in order to illustrate the dependence of this feature on the strength of the gravitational field.

\subsection{Equatorial radius and mass}
\label{sec:radius}

Although $C_e$ is roughly constant along constant-central-density spin-up sequences, neither $M$ nor $R_e$ is constant along these sequences. Typical changes in the mass or radius are at least an order of magnitude larger than the change in their ratio. 

{Other authors have examined the universality in the increase in mass for some special cases. \citet{2016MNRAS.459..646Breu} calculated the values of mass, called the critical mass, along the quasi-radial instability line (the thin dashed line shown in Figure \ref{fig:m_r_nu_Pol0}).  They found a universal dependence of the critical masses normalized by the TOV mass on the dimensionless spin parameter $\chi$. 
\citet{2020PhRvD.101f3029Shao} added to this relation the dependence of the critical mass on the compactness of the TOV model, as well as a universal relation for the radii of the critical models.
These relations for the critical masses were shown to have some EOS dependence when phase transitions and differential rotation are included in the computations \citep{2019EPJA...55..149Bozzola}. In this paper, we examine a more general universality in the increase in mass (and radius) for all uniformly rotating neutron stars, not just the critical stars.}

\begin{figure}
    \centering
    \plottwo{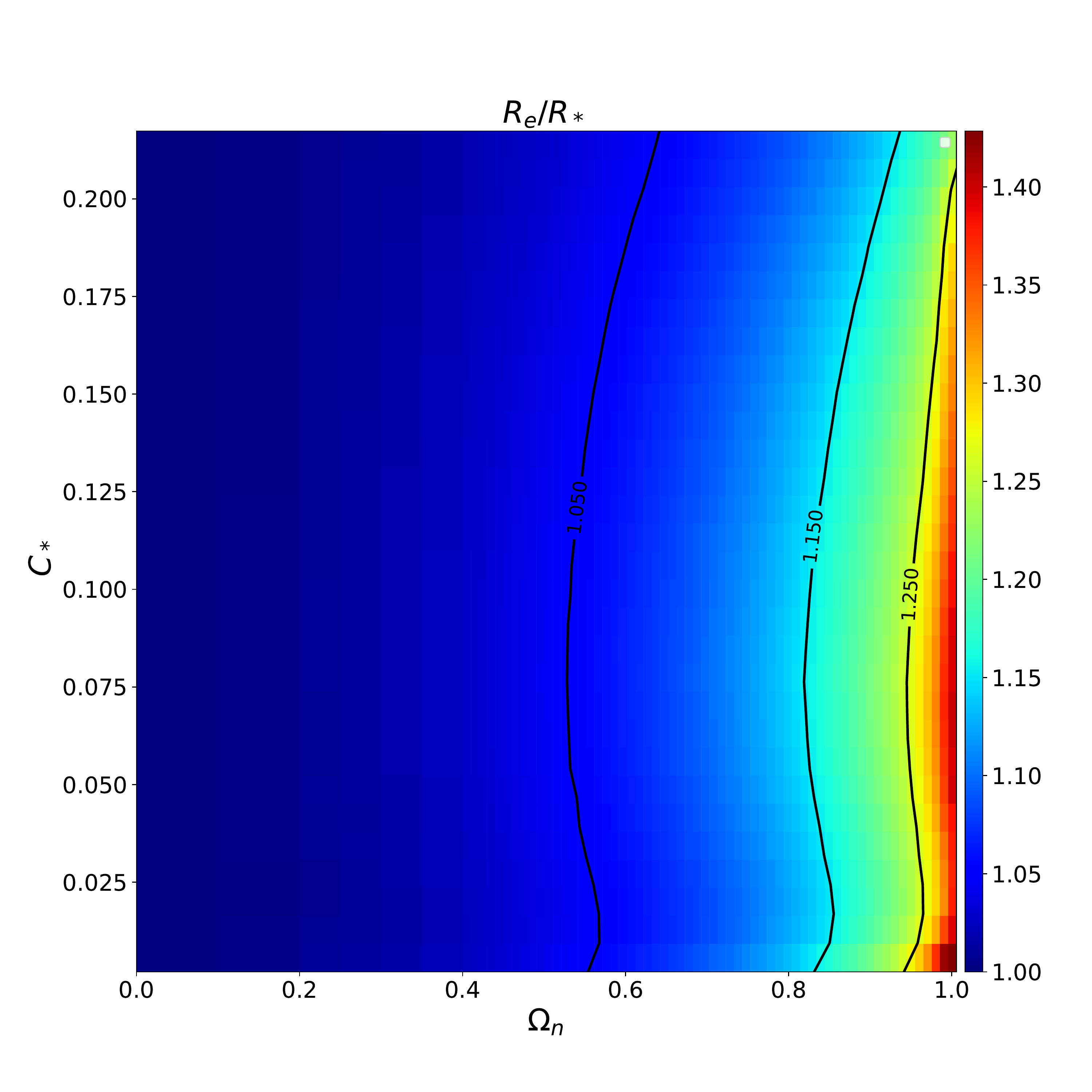}{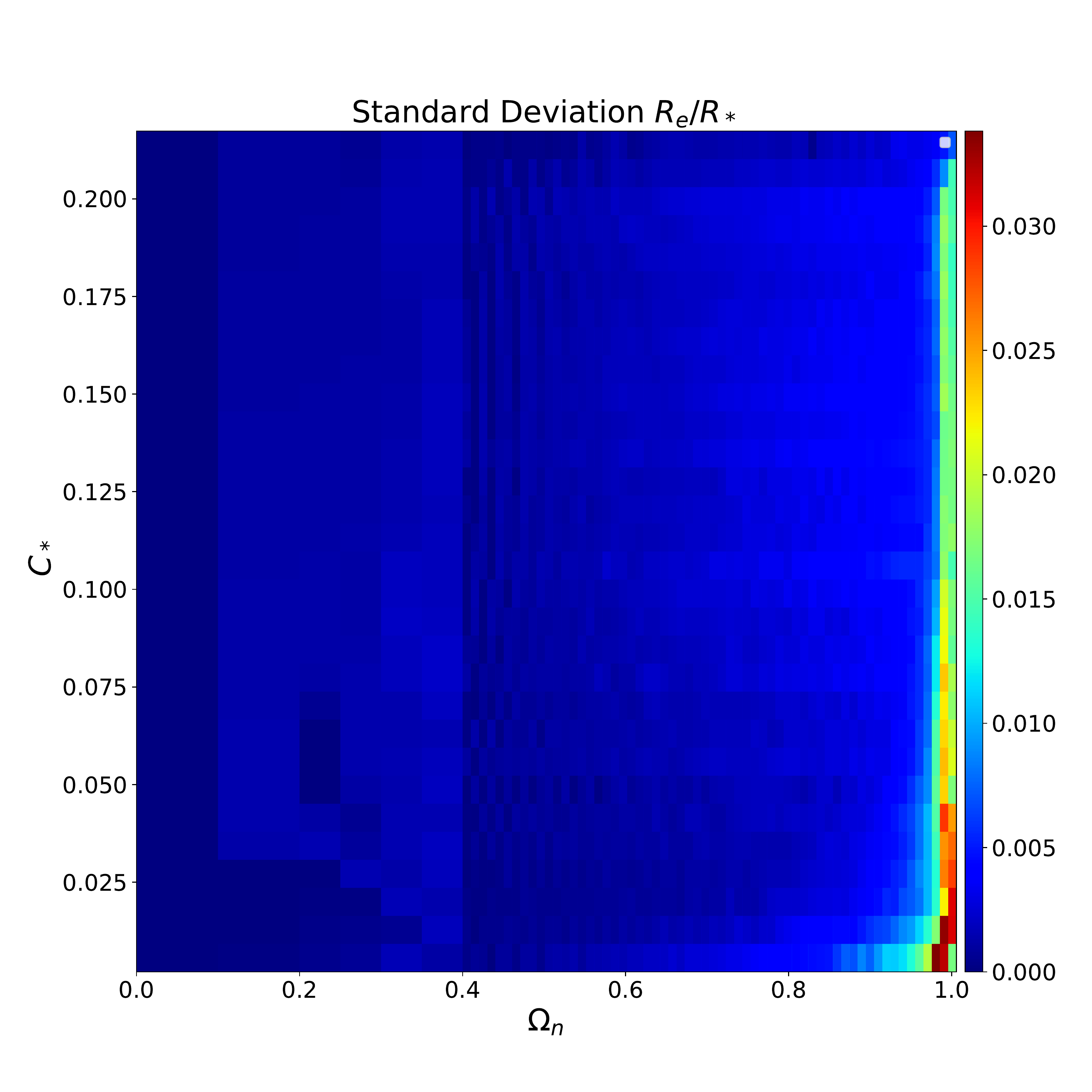}
    
    \caption{Average (left) and standard deviation (right)  of the increase in equatorial radius due to rotation as a function of $\Omega_n$ and $C_*$. The colour bar represents the average (or standard deviation) of the value of $R_e/R_*$ for all 32 EOS in each bin.  }
    \label{fig:radius-increase}
\end{figure}

\begin{figure}
    \centering
    \plottwo{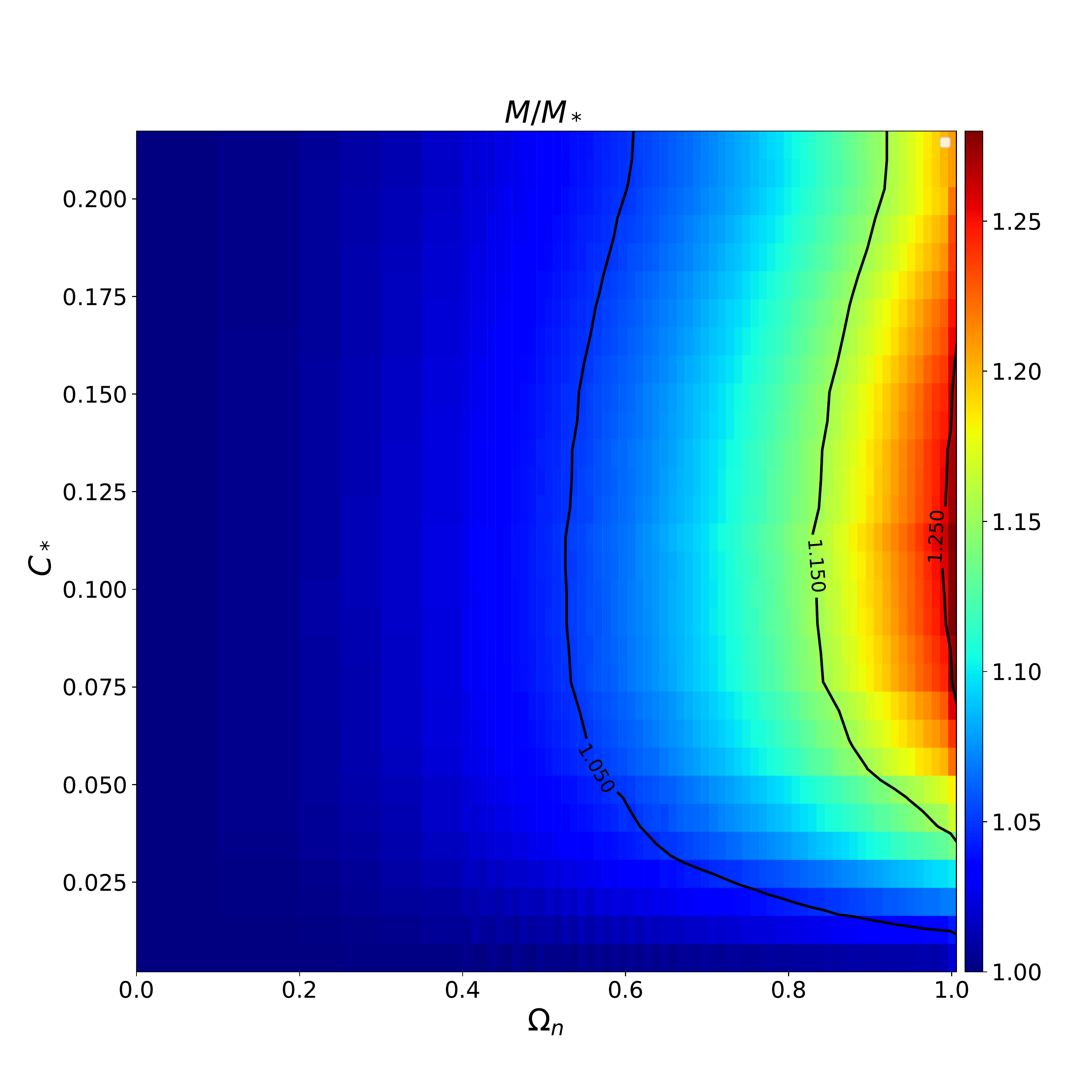}
    {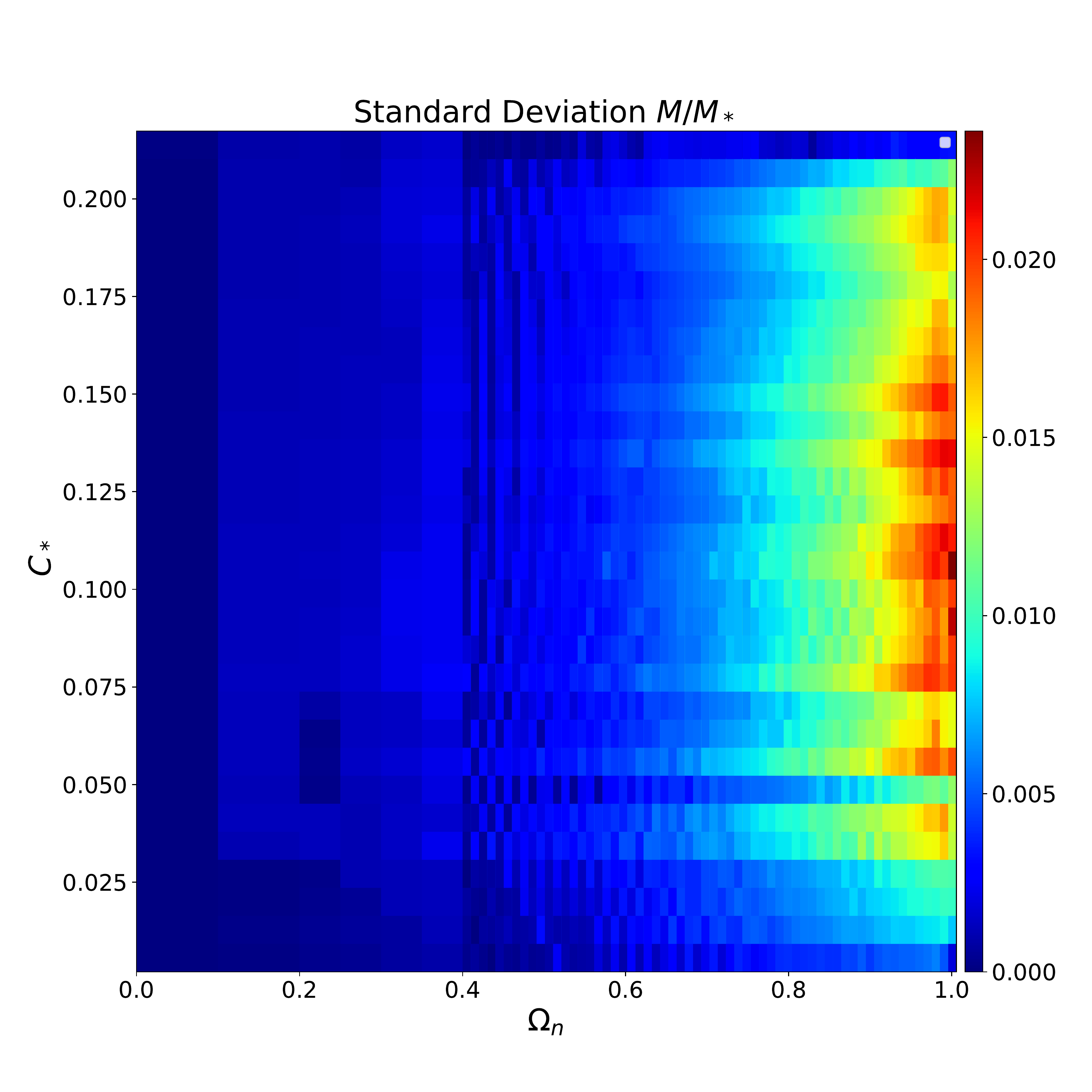}
    \caption{Average (left) and standard deviation (right) in the increase in mass due to rotation as a function of $
    \Omega_n$ and $C_*$. The colour bar represents the average (or standard deviation) of the value of $M/M_*$ for all 32 EOS.  }
    \label{fig:mass-increase}
\end{figure}

The rotationally-induced increase in the neutron star's  equatorial radius and mass is shown in Figures \ref{fig:radius-increase} and \ref{fig:mass-increase} respectively. The average values and standard deviations of $R_e/R_*$ and $M/M_*$  (over all 32 EOS) are shown in each bin of constant $C_*$ and $\Omega_n$. Figure \ref{fig:radius-increase} shows that on average (left panel), the radius increases by about 15\% when the neutron star is rotating at about 80\% of the Kepler limit. The standard deviation in the radius increase is less than 1\% of $R_*$ for the same range. Since the increase in radius is almost independent of the EOS, and only depends on $C_*$ and $\Omega_n$, the increase appears to be  ``universal".

Similarly, the average and standard deviation in the increase in the mass due to rotation is shown in Figure \ref{fig:mass-increase}.
The mass increases by 15\% at a somewhat larger normalized angular velocity, and there is more dependence on compactness than is seen in Figure \ref{fig:radius-increase} for the increase in radius.
The standard deviation in $M/M_*$ grows as large as 1\% at spins of 80\% of the Kepler limit. While this is a larger variation than seen in the increase in radius, this is still almost independent of the EOS.

Based on these results, we made the  hypothesis that the changes in mass and equatorial radius along a constant-density spin-up sequence can be approximated by a surface in a space spanned by $C_*$ and $\Omega_n$. 
The best-fit equations for the surfaces representing these quantities are
\begin{eqnarray}
\frac{R_e}{R_*} &=&  1 +  ( e^{A_r \Omega_n^2} - 1 + B_r \left[ \ln( 1 - (\frac{\Omega_n}{1.1})^4)\right]^2)  
\times \left( 1 + \sum_{i=1}^{5} a_{r,i} C_*^i \right) \label{eq:Rfit}\\
\frac{M}{M_*} &=& 1 + ( e^{A_m \Omega_n^2} - 1)  \times \left( \sum_{i=0}^{4} a_{m,i} C_*^i \right),\label{eq:Mfit}
\end{eqnarray}
where the best-fit coefficients and the $R^2$ statistics for the fits are shown in Table \ref{tab:coeff}. {The functional forms of these equations were found by first plotting $R_e$ and $M$ as surfaces depending  on $C_*$ and $\Omega_n^2$ and noting the approximately exponential dependence on the square of the angular velocity by eye. The logarithmic term was added to model the rapid change in radius as the Kepler limit is approached. The factor of 1.1 was required since the empirical nature of the fits leads to some stars having values of $\Omega_n>1.$ }

In order to quantify the quality of these best-fit equations, we introduce the absolute deviation of some quantity $Q$. The absolute deviation of $Q$, $Dev(Q)$ for a particular rotating neutron star model $Q_i$ from the best-fit surface $Q_{fit}$ is defined by
\begin{equation}
    Dev(Q) = \frac{|Q_i-Q_{fit}|}{Q_i},
\end{equation}
which is the fractional error incurred by using the best-fit function. Examination of the absolute deviations is useful for flagging possible outlier EOS that are not described well by the best-fit functions.

%%%

\begin{figure}[h]
    \centering
    \plottwo{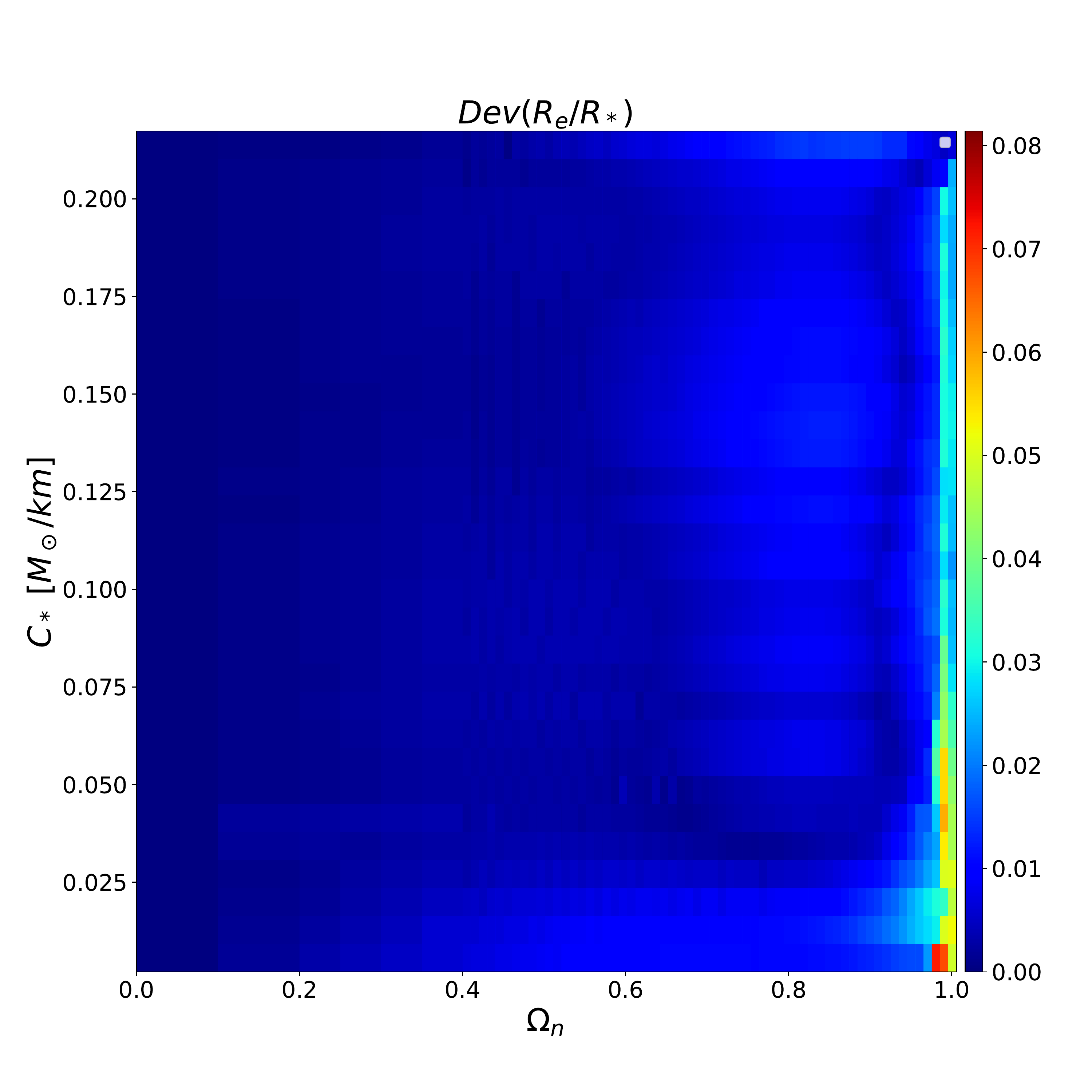}{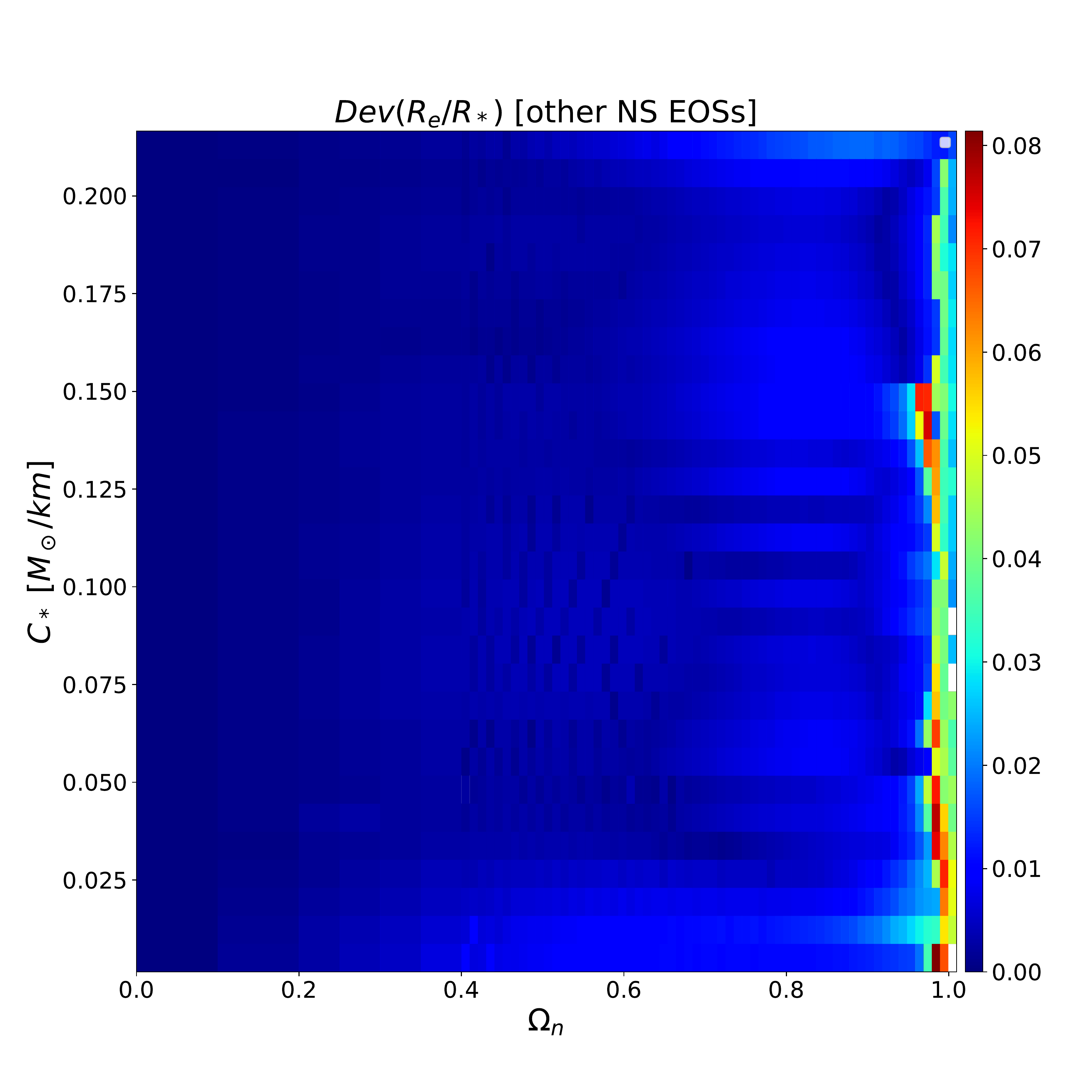}
\plottwo{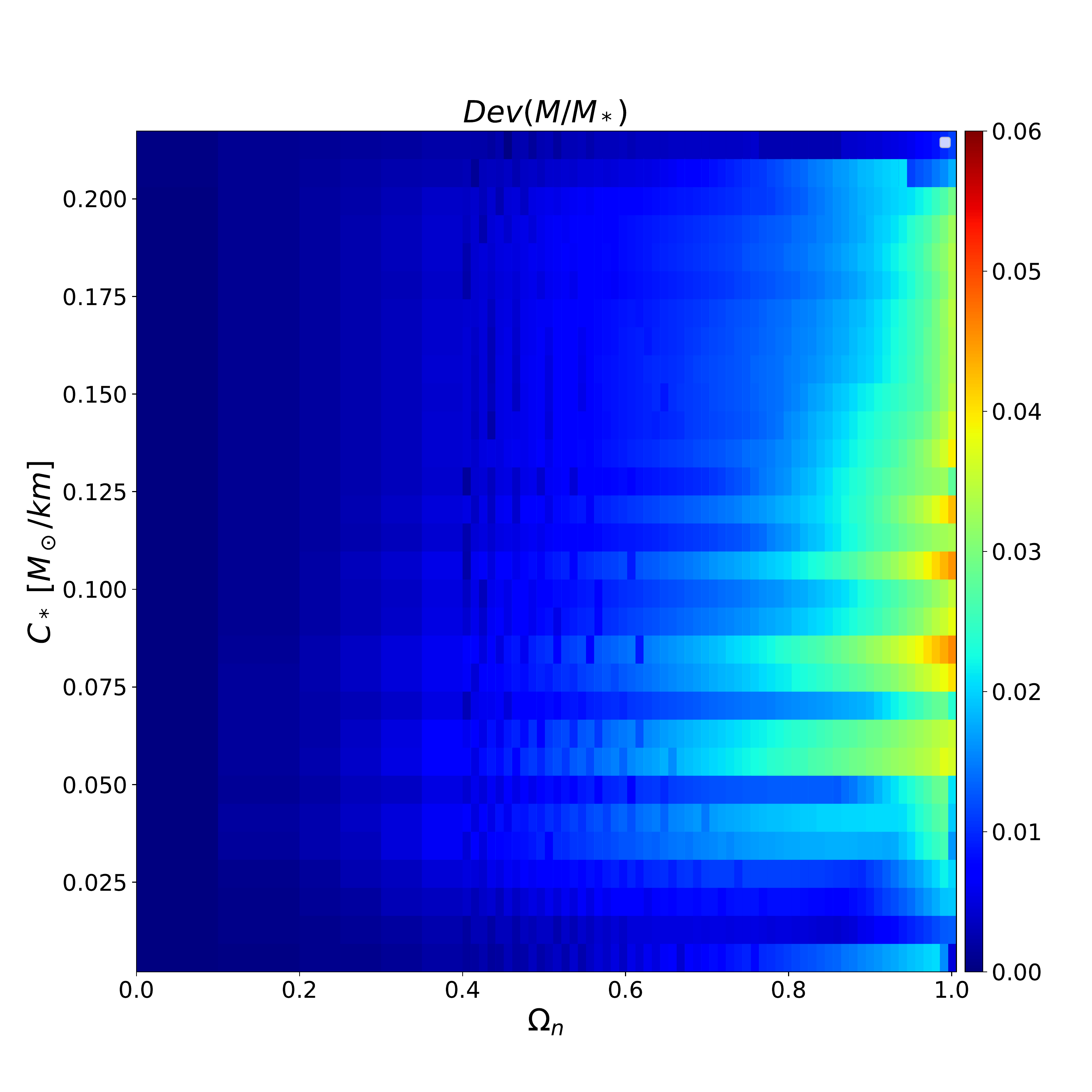}{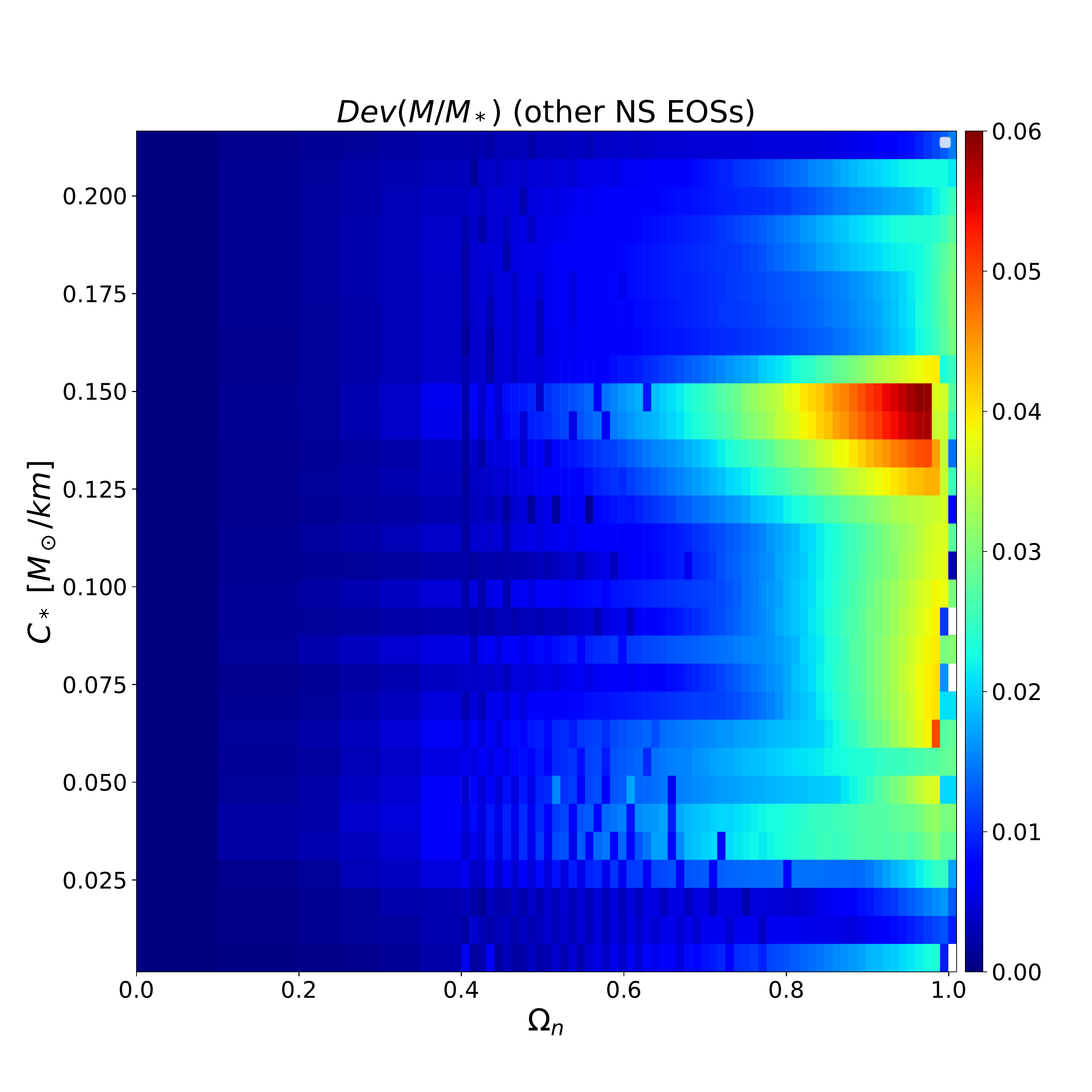}
        \caption{Largest absolute deviation in the radius (upper) and mass (lower). 
        The colour in each $C_*$-$\Omega_n$ bin represents the largest absolute deviation in the ratio $R_e/R_*$ or $M/M_*$ for any of the 32 randomly generated EOS (left) or tabulated hadronic or hybrid EOS (right). }  
    \label{fig:DevRadius}
\end{figure}

The largest absolute deviations introduced by the best-fit value of $R_e/R_*$, Equation (\ref{eq:Rfit}) for any of the
EOS are shown in Figure \ref{fig:DevRadius} (upper panels). The colour in each pixel represents the largest absolute deviation of the fitting function for the scaled equatorial radius $R/R_*$, from the value computed for any star with the given value of $C_*$ and $\Omega_n$. The left-hand panel shows the largest absolute deviation within the 32 randomly generated EOS while the right-hand panel shows the largest deviation for the smaller set of 
tabulated hadronic and hybrid EOS described in the last paragraph of Section \ref{sec:eos}. 
Similarly, the largest absolute deviation of $M/M_*$ for the same groups of EOS is shown in the lower panels of \ref{fig:DevRadius}.

Figure \ref{fig:DevRadius} shows that for values of $\Omega_n \le 0.8$ the error introduced by the formula for the rotating star's radius given by Equation (\ref{eq:Rfit}) is typically less than 2\%  hadronic and hybrid EOS. The deviations for $M/M_*$ are larger than the radius deviations, and more equation of state dependence can be seen in this Figure. Stars with $\Omega_n = 0.8$ can have deviations in the rotating mass that are as large as 4\%, depending on the value of $C_*$ and the EOS. The largest deviations come about mainly from the hyperonic EOS, which has a very low maximum mass of 1.5 $M_\odot$.
Bare quark stars show a larger deviation, as large as 5\% in mass at 60\% of Kepler. As has been seen for other universal relations, the bare quark EOS probably should be treated separately from hadronic EOS. We speculate that the larger EOS dependence seen in the $M/M_*$ plots (for all EOS) is due to some dependence on how close the star's mass is to the maximum mass.

The absolute deviations in the best-fit formulae can rise to about 8\% error for stars rotating close to the Kepler limit. This is not surprising, since it has been seen that other universal relations, for instance the I-Love-Q relations break down for very rapid rotation \citep{2014ApJ...781L...6Doneva}.

\subsection{Approximate universality}

The approximately constant value for the equatorial compactness on a constant-density spin-up sequence can be understood intuitively through a simple Newtonian model. In Euclidean space, the volume of an oblate spheroid with equatorial radius $R_e$ and polar radius $R_p$ is $V = 4\pi R_e^2 R_p/3$. Suppose that this ellipsoid is a member of the same sequence of models as a non-rotating sphere with radius $R_*$ and volume $V_*$, with $R_e = R_* + \Delta R_e$ and $R_p = R_* - \Delta R_p$. For small changes in radius, the change in volume is
\begin{equation}
    \Delta V = V_* \frac{R_e}{R_*} \left(2 \frac{\Delta R_e}{R*} - \frac{\Delta R_p}{R*} \right) 
    \simeq V_* \frac{\Delta R_e}{R_*},
\end{equation}
if $\Delta R_p \simeq \Delta R_e$. 

In the case of a uniform density star, if the central density is kept constant, then the logarithmic change in the total mass of the star will increase at the same rate as the logarithmic change in the volume, so 
\begin{equation}
    \frac{\Delta M}{M_*} = \frac{\Delta V}{V_*}.
    \label{eq:mass-volume}
\end{equation}
In the situations where $\Delta R_p \simeq \Delta R_e$, this leads to 
the logarithmic change in the mass being roughly equal to the logarithmic change in the equatorial radius, leading to constant compactness along constant central-density sequences. 

In reality, the density is not constant inside a neutron star. However, in the case of Newtonian polytropes, the average density of a star is equal to a constant (for a particular EOS) times the central density. If this property is preserved as a star spins faster, then a sequence with constant central-density will also have constant average density, leading to equation
(\ref{eq:mass-volume}). 

\begin{figure}[h]
    \centering
    \includegraphics[width=90mm]{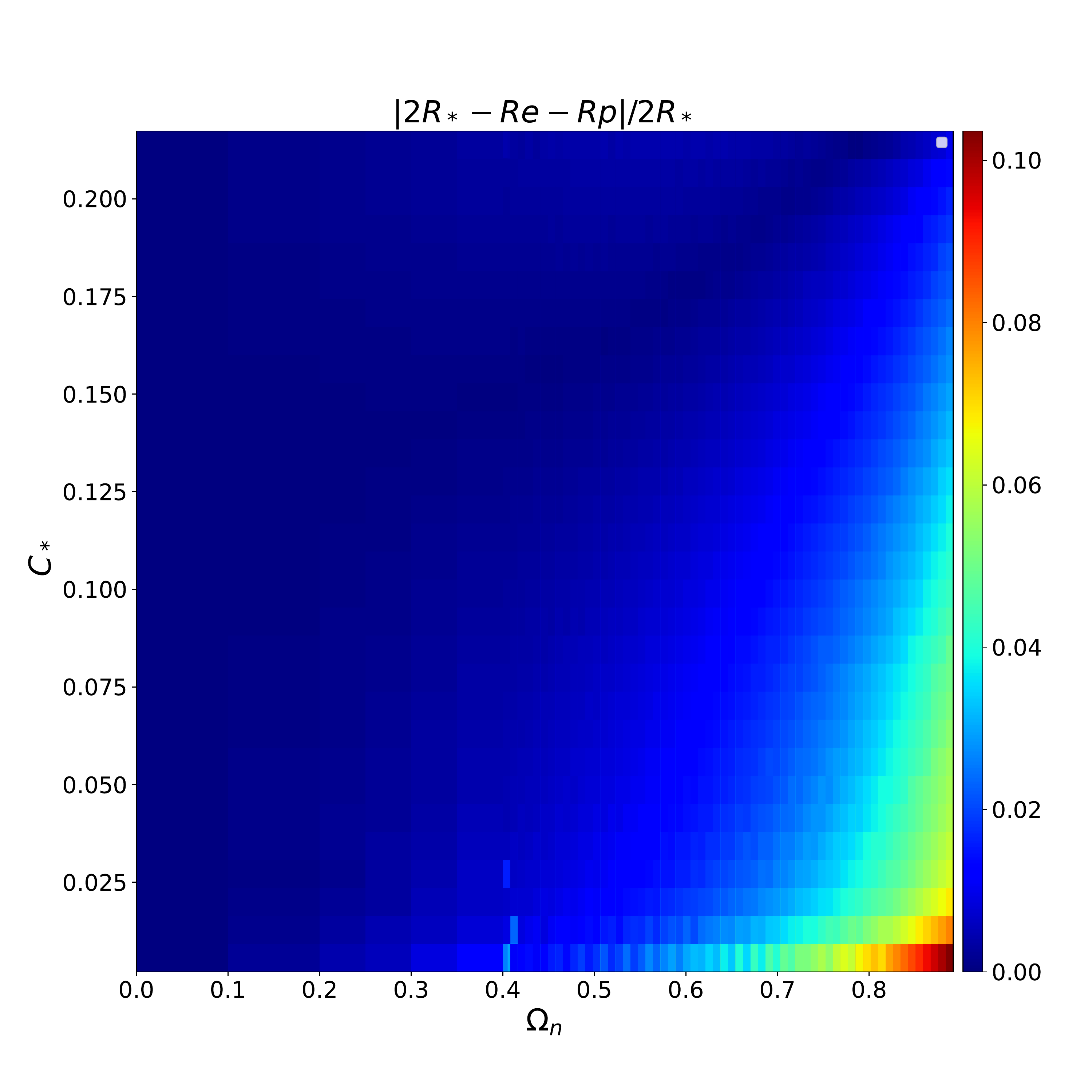}
    \caption{Neutron star deformation asymmetry. The colour scale shows the largest absolute value (over the 32 EOS) of the dimensionless rotational deformation asymmetry, $(2R_*-R_e-R_p)/(2R_*)$, for each bin of $C_*$ and $\Omega_n$.  }
    \label{fig:deformation}
\end{figure}

In order to quantitatively show how the polar radius of a star changes relative to the equatorial radius, we introduce a quantity, $\alpha$, that measures the asymmetry in the rotational deformation, defined by
\begin{equation}
    \alpha = \frac{2 R_* - R_e - R_p}{2R_*}.
\end{equation}
The rotational deformation asymmetry, $\alpha$, vanishes for zero rotation by definition. Rotating stars with $\Delta R_e = \Delta R_p$ will also have vanishing $\alpha$. Note that a star can be highly elliptical even if $\alpha$ vanishes. The definition of $\alpha$ is not a coordinate-independent definition. In our application, we compute  $\alpha$ using the Schwarzschild-like radial coordinate, since it is most closely connected to physically meaningful quantities like the circumference.  Figure \ref{fig:deformation} shows the largest absolute value of $\alpha$ for each value of $C_*$ and $\Omega_n$. The values of $\alpha$ are small compared to the values of $\Delta R_e/R_*$ shown in Figure \ref{fig:radius-increase}, and trends to smaller values with increasing compactness. The vanishing of the rotational deformation asymmetry in the limit of very compact stars is most likely related to the iso-density self-similarity discussed by \citet{2017PhR...681....1Yagi}.

When we examine the constant-density contours inside a rotating neutron star, we see a general qualitative trend that the deformation of the star is dominated by the deformation of the crust. This suggests a simple toy model of a dense, incompressible core that stays roughly spherical and a low density crust that is easily deformed by the centrifugal force. If we think of the two regions as two springs in series, with a stiff spring attached to a wall representing the core, and a soft spring representing the crust, the overall effective spring constant will have a value similar to the soft spring's constant (in the limit that the crust's spring constant is much smaller than the core's spring constant). The applied centrifugal force, proportional to $\Omega_n$, then induces a strain in the crust that depends on $\Omega_n$, but is almost independent of the length of the spring (or thickness of the crust). Since most of the EOS in our libraries (including the tabulated hadronic and hybrid EOS) have similar crust EOS, it is natural that the change in equatorial radius with spin should show the almost universal behaviour, as seen in Figure \ref{fig:radius-increase}. The bare quark star EOS does not include a crust, and with this toy model it is not surprising that bare quarks stars are not as well described by universal relations.

\subsection{Application to EOS inference}
\label{sec:inference}

The empirical relations for the mass and radius of a rotating neutron star given the 
values found from the spherically symmetric TOV equations with the same central density lead to a simple set of corrections that can be used in EOS inference codes. 

We imagine that a number of neutron stars' masses, equatorial radii, and spins have been measured, for instance using pulse-profile modelling as is done by NICER. If we have $N$ measurements, the $i$-th neutron star has  spin frequency $\nu_i$, and the mass and equatorial radius will be described by a probability distribution. The spin-corrected EOS inference procedure is to first choose an EOS, and then solve the TOV equations for a range of central densities to construct a zero-spin mass-radius curve. For each value of central density use the zero-spin values $M_*$, $R_*$, and the spin frequency of the $i$th star to compute the predicted mass and radius using equations (\ref{eq:Mfit}) and (\ref{eq:Rfit}). The likelihood of this empirically predicted mass and radius can then be computed using the $i$th mass-radius distribution, and then the procedure repeated for other central densities until the central density that maximizes the $i$th likelihood function has been found (for this specific choice of EOS). Since the empirical relations are simple algebraic expressions, they are computationally inexpensive to implement. This makes it simple to add rotation to an EOS inference code that simply solves the TOV equations. 

Whether or not these empirical corrections to the rotating star's mass and radius should be added to an EOS inference code depends on the precision of the measurements, and the relative increase in the star's mass and radius due to rotation. For instance, NICER measured the mass and radius of the pulsar J0030 with a precision of about 10\% in both quantities \citep{2019Miller, 2019Riley}. This pulsar's spin frequency is 205 Hz so comparing with values in Table \ref{tab:typicalvalues}, the increases in mass and radius are less than 1\% for typical mass and radius values. 
The small size of these increases means that the simple use of the TOV equations in the EOS inference results of \citet{2019Miller} and \citet{2019Raaijmakers} is valid. It is only when the precision of mass and radius parameter estimation for the NICER pulsars is similar to the typical magnitudes of the changes given in Table \ref{tab:typicalvalues} will the rotational corrections presented in this paper be significant.

\section{Inverse mapping}
\label{sec:inverse}

Given that we have found a mapping between the mass and radius of non-rotating neutron stars to rotating neutron stars, it is natural to look at inverse mappings. The inverse mapping is potentially useful if many neutron stars, with different spins, have precise mass and equatorial radius measurements. An inverse mapping allows the construction of an effective zero-spin mass-radius relation using multiple observations.

\begin{figure}[h]
    \centering
    \includegraphics[width=0.5\textwidth]{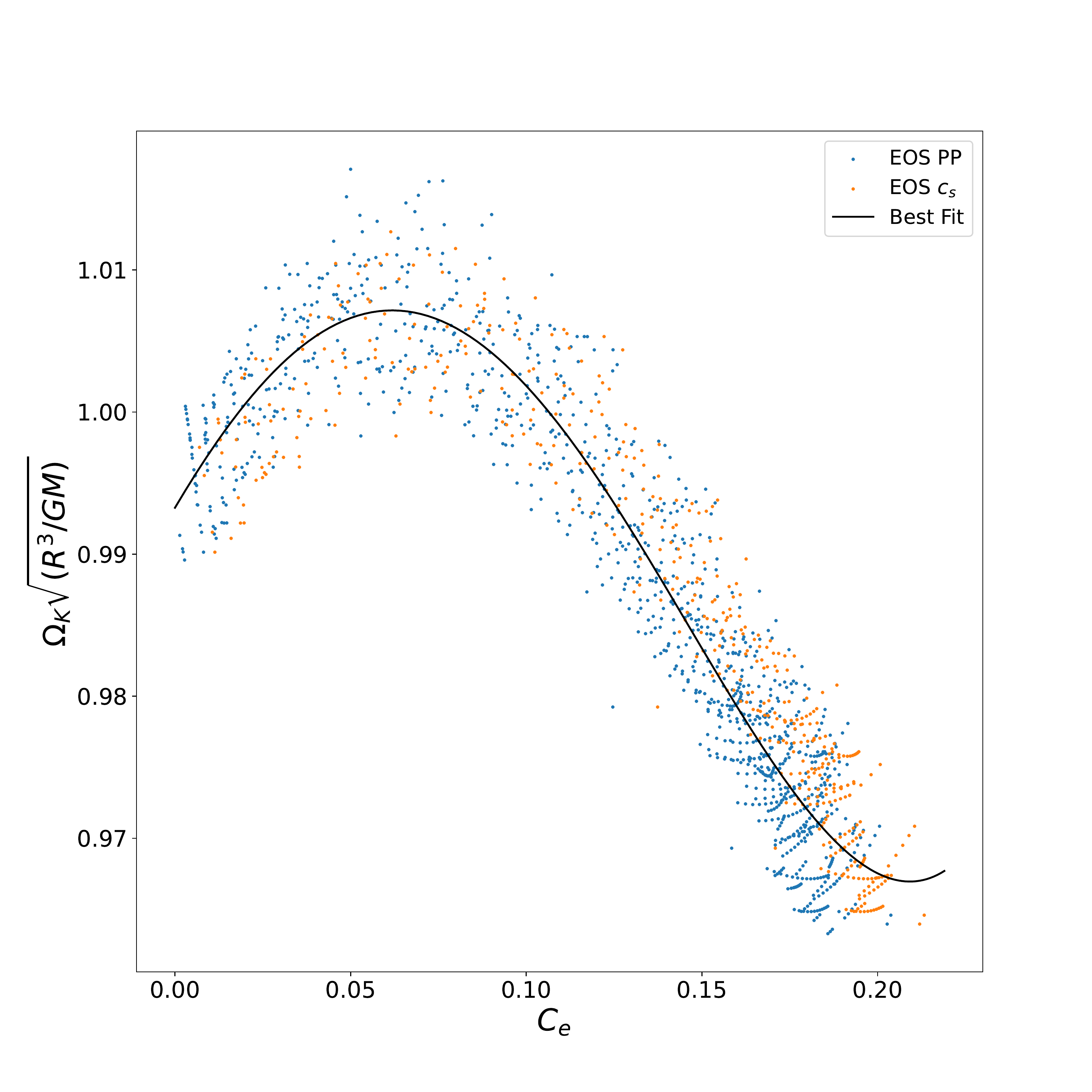}
    \caption{The normalized Kepler limit angular velocity versus the compactness ratio of the rotating neutron star. Blue and orange points represent neutron stars spinning at the Kepler limit computed with either a PP or CS EOS respectively. The best-fit equation (\ref{eq:kepler-inv}) to these points is shown as a solid curve.}
    \label{fig:kepler-inv}
\end{figure}

The inverse map from the measurement of a rotating star's mass and radius to a non-rotating star with the same central energy density can be constructed using a method similar to Section \ref{sec:universal}. The first step is to construct a new normalized angular velocity, since the previous definition relies on knowledge of $M_*$ and $R_*$. A natural choice is to compute the Kepler angular velocity normalized by $\sqrt{GM/R_e^3}$, where $M$ and $R_e$ are the mass and equatorial radius of a neutron star rotating at the Kepler limit. Figure \ref{fig:kepler-inv} shows this normalized Kepler frequency plotted versus the equatorial compactness for all 32 EOS. Each blue or orange point represents a maximally rotating star from one of the constant density sequences computed using either a PP or CS EOS. The solid curve shows a best-fit curve for these points, which remarkably, is within a couple percent of 1. The equation of this curve is
\begin{equation}
    \Omega_{K2} = \sqrt{\frac{GM}{R_e^3}} \times \sum_{i=0}^{4} b_i C_e^i.
    \label{eq:kepler-inv}
\end{equation}
The best-fit coefficients for this equation can be found in Table \ref{tab:coeff}.

The Kepler angular velocity is, by definition, both the angular velocity of the star spinning at the mass-shed limit and the angular velocity of a particle orbiting  the star at the equator (in the same direction as the star's rotation). The orbital angular velocity of a particle orbiting a non-rotating star is  $\Omega_{s,orb}=\sqrt{GM/R^3}$. Since the formula for the orbital frequency at the mass-shed limit is almost the same, this suggests that the orbital frequency at intermediate spins be given by a similar formula $\Omega_{orb}(\Omega) = F \sqrt{GM/R_e^3}$, where $F\sim 1$. When we plot (not shown) the orbital frequencies for all models, we find that while there is more scatter, the constant is always within 3\% of 1. This does not depend on the stability of the orbit. We have not attempted to fit this relationship for the orbital angular velocity for intermediate spin rates, but note that approximating $F\sim 1$ may be good enough for other applications.

{The spin parameter $\chi$ is approximately proportional to the star's angular velocity. In the case of stars rotating at the Kepler frequency, it has been argued \citep{2016MNRAS.459..646Breu,2020PhRvD.101f3029Shao} that since $\Omega_K \sim \sqrt{GM/R^3}$, the value of the spin parameter at the Kepler limit should be $\chi_K \sim I/(MR^2) \times \sqrt{Rc^2/GM}$, where $I$ is the moment of inertia, and $I, M,$ and $R$ are all evaluated at the Kepler limit. When combined with the universal relations for $I$ \citep{2020PhRvC.101a5805Koliogiannis} at the Kepler limit, \citet{2020PhRvD.101f3029Shao}
showed that this leads to a very simple formula for the largest value of $\chi$ allowed by an equation of state, $\chi_{max} = 1.24 (\sqrt{GM/Rc^2})_{TOV}$. This leads to a prediction of about a 10\% variation in the maximum possible value of $\chi$, depending on the EOS. It would be interesting to investigate the dependence of mass and radius on dimensionless parameters such as $\chi/\chi_K$. However, our preliminary investigations found that the dependence on $\chi/\chi_K$ introduced more scatter than occurs with a quantity such as $\Omega/\Omega_{K2}$ that traces the relative magnitudes of the centrifugal and gravitational forces. }

Given an observation of a rotating neutron star's mass, equatorial radius, and angular velocity, we can compute the quantities $C_e$ and a normalized angular velocity
\begin{equation}
    \Omega_{n2} = \frac{\Omega}{\Omega_{K2}}.
\end{equation}
Based on our library of neutron star models, we find the following best-fit equations for the mass and radius of a non-rotating neutron star with the same central density as the rotating star,
\begin{eqnarray}
\frac{R_e}{R_*} &=&  1 +  ( e^{A_{2} \Omega_n^2} - 1 + B_{2} \left[ \ln( 1 - (\frac{\Omega_{n2}}{1.1})^4)\right]^2)  
\times \left( 1 + \sum_{i=1}^{5} b_{r,i} C_e^i \right) \label{eq:Rinvfit}\\
\frac{M}{M_*} &=& 1 + \left( \sum_{i=1}^4 d_i \Omega_{n2}^i\right)  \times \left( \sum_{i=1}^{4} b_{m,i} C_e^i \right),\label{eq:Minvfit}
\end{eqnarray}
where the coefficients appearing in these equations can be found in Table \ref{tab:coeff}.
If we have a few neutron star mass-radius measurements, these equations can be used to produce an effective zero-spin mass-radius curve for this group of neutron stars.

Figure \ref{fig:m_r_nu_Pol0} demonstrates the use of the inverse map on the example EOS PP0. Each blue dot representing a rotating neutron star has been mapped to a non-rotating neutron star with the same central density,  represented with an orange dot, using Equations (\ref{eq:kepler-inv}) - (\ref{eq:Minvfit}). If the inverse map were perfect, all of the orange dots for a particular central density would overlap and lie exactly on the solid black TOV curve. It can be seen in Figure \ref{fig:m_r_nu_Pol0} that there is a small spread in the orange dots, however, the spread is much smaller than the typical uncertainty in the measurements.

NICER observations of PSR J0030+0451, which spins with a frequency of $\nu = 205$ Hz, were analyzed by \cite{2019Miller} and \cite{2019Riley} leading to estimates for this pulsar's mass and equatorial radius with precisions (at one sigma) of about 10\%. 
For example,
\cite{2019Miller} found values M = $1.44^{+0.15}_{-0.14}$ $M_\odot$ and $R_e$ = $13.02^{+1.24}_{-1.06}$ km for the rotating star. Making use of Equations (\ref{eq:kepler-inv}) - (\ref{eq:Minvfit}) with $M=1.44 M_\odot$, $R_e=13.02$ km, and $\nu=200$ Hz, we find that the equivalent non-rotating mass and radius are $M_*=1.43 M_\odot$ and $R_*=12.94$ km. These changes in mass and radius due to rotation are about an order of magnitude smaller than the measurement precision, so there is no loss of accuracy if the rotational correction terms are ignored.

The precision of the mass and radius measurements made by NICER will improve as more data is accumulated with longer observations. However, NICER's target precision of 5\% for the radius measurement is still larger than the $~1\%$ typical change in radius predicted for neutron stars spinning at the relatively slow rate of 200 Hz. The other pulsars observed by NICER spin at similar rates, so the order of magnitudes for their spin corrections are similar. However, future missions, such as Strobe-X \citep{2019arXiv190303035Ray} and eXTP 
\citep{2019Watts}
will have a 10x larger collecting area, which will allow for more precise radius measurements as well as observations of fainter pulsars.  In addition, these missions are likely to target accreting ms-period X-ray pulsars that spin more rapidly.  The universal formulae for the changes in mass and radius will be more important for these future X-ray missions.

\section{Discussion}
\label{sec:discussion}

Neutron star properties such as their masses and radii are decidedly non-universal, since they depend strongly on the EOS of supranuclear density matter. However, given a value of a neutron star's mass and radius, many secondary properties only appear to depend on the given mass and radius, and are almost independent of the choice of EOS. In this paper we investigate the universality of the change in mass and radius caused by rapid rotation.

We computed equilibrium sequences of rapidly rotating neutron stars with constant central density to explore the universality of the increase in mass and equatorial radius. We found that the fractional increase of mass and radius can be fit with an empirical formula that depends only on the mass and radius of the non-rotating star in the sequence and the rotating star's spin frequency. As part of this method, we have also derived a formula for the fastest spin frequency of a neutron star given the mass and radius of a non-rotating neutron star with the same central density. These empirical formulae can then be applied as corrections to EOS inference codes, since EOS inference codes compute the mass and radius of a non-rotating star with a given central density. 

We also explored the surprising property that the equatorial compactness of a neutron star is very close to constant along a constant central-density sequence. This property appears to be related to an internal symmetry that keeps the relative decrease in the polar radius close to equal in magnitude to the increase in the equatorial radius. 

It is also interesting to consider whether an effective zero-spin mass-radius relation can be constructed from observations of the masses and radii of many neutron stars with a wide variety of spin frequencies. We constructed a set of equations that allow the mapping of a rotating star's mass and radius to a corresponding non-rotating neutron star with the same central density.

The rotational corrections derived in this paper are generally smaller than the typical uncertainties in determining the masses and radii of rotating neutron stars. So at present, these correction factors are not required in EOS inference codes used with results obtained from pulse-profile modelling done by NICER. However, we expect these formulae to be useful in the future when more precise measurements are possible, and when radius measurements of more rapidly rotating neutron stars are possible. 

\section{Acknowledgements}

We thank Jorge Calderon Noguez, Rodrigo Fernandez, and Craig Heinke for helpful discussions, and the anonymous referee for useful comments. This research was supported by a grant from NSERC to SMM.

\bibliography{biblio}{}
\bibliographystyle{aasjournal}

%% This command is needed to show the entire author+affiliation list when
%% the collaboration and author truncation commands are used.  It has to
%% go at the end of the manuscript.
%\allauthors

%% Include this line if you are using the \added, \replaced, \deleted
%% commands to see a summary list of all changes at the end of the article.
\listofchanges

\end{document}